\documentclass[sigconf]{acmart}

\usepackage{enumitem}
\usepackage{hyperref}
\usepackage{tabularx}
\usepackage{xurl}  
\usepackage{graphicx}
\usepackage{amsmath}
\usepackage{hyperref}
\usepackage{multirow, multicol, makecell}
\usepackage[breakable,skins]{tcolorbox}
\usepackage{enumitem}

\title{“TODO: Fix the Mess Gemini Created”: Towards Understanding GenAI-Induced Self-Admitted Technical Debt}

\newcommand{\name}{\textbf{GenAI-Induced Self-admitted Technical debt}}
\newcommand{\nameacronym}{\textbf{GIST}}

\author{Abdullah Al Mujahid}
\affiliation{%
  \institution{Missouri University of Science and Technology}
  \department{Department of Computer Science}
  \city{Rolla}
  \state{MO}
  \country{USA}
}
\email{amgzc@mst.edu}

\author{Mia Mohammad Imran}
\affiliation{%
  \institution{Missouri University of Science and Technology}
  \department{Department of Computer Science}
  \city{Rolla}
  \state{MO}
  \country{USA}
}
\email{imranm@mst.edu}



\begin{document}

\begin{abstract}
As large language models (LLMs) such as ChatGPT, Copilot, Claude, and Gemini become integrated into software development workflows, developers increasingly leave traces of AI involvement in their code comments. Among these, some comments explicitly acknowledge both the use of generative AI and the presence of technical shortcomings.

Analyzing 6,540 LLM-referencing code comments from public Python and JavaScript-based GitHub repositories (November 2022–July 2025), we identified 81 that also self-admit technical debt (SATD). 
Developers most often describe postponed testing, incomplete adaptation, and limited understanding of AI-generated code, suggesting that AI assistance affects both when and why technical debt emerges.
We term {\name} {(\nameacronym{})} as a proposed conceptual lens to describe recurring cases where developers incorporate AI-generated code while explicitly expressing uncertainty about its behavior or correctness. 
\end{abstract}

\maketitle

\begin{sloppypar}

\section{Introduction}
Large Language Models (LLMs) are rapidly transforming software development workflows. Integrated tools such as GitHub Copilot, ChatGPT, and Gemini now assist with code generation, refactoring, documentation, and testing. While these systems increase productivity, they also introduce novel forms of risk and uncertainty. Developers must evaluate, correct, and maintain code partially written by non-human agents.

\begin{figure}[tb]
  \centering
  \includegraphics[width=\columnwidth]{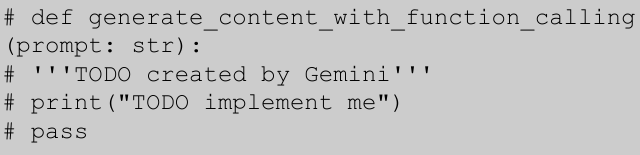}
  \caption{Example self-admitted comment illustrating requirement debt}
  \label{fig:req-comment}
\end{figure}

Code comments, particularly those that self-admit technical shortcomings, offer a unique window into how developers interpret and manage AI-assisted contributions. In software engineering, self-admitted technical debt (SATD) signals areas requiring future improvement. When such comments reference LLMs, they reveal not only perceived code deficiencies but also evolving conceptions of accountability between human and machine collaborators.
Figure~\ref{fig:req-comment} shows one such example, where a developer notes that a function was generated by Gemini and remains unfinished, illustrating how AI involvement and deferred implementation converge to create a potential debt.

While prior research has established taxonomies and automated methods for identifying SATD~\cite{maldonado2015detecting, alves2014Towards, rantala2020prevalence, rantala2024keyword, alves2016identification}, little is known about how technical debt manifests in code influenced by generative AI. Early studies on AI-assisted development highlight productivity benefits~\cite{barke2023grounded,coutinho2024role}, but also point to challenges of correctness, explainability, and trust. What remains unclear is how developers themselves recognize and document these challenges in practice, and whether AI involvement reshapes the kinds of debt that are self-admitted during development.

This study explores SATD in the context of generative AI-assisted software development. By analyzing code comments that explicitly reference both LLM use and technical debt markers, we investigate how developers identify and articulate shortcomings in AI-generated code. Specifically, we ask:

\noindent
{\bf RQ1: What types of self-admitted technical debt emerge in developer comments that acknowledge both LLM involvement and technical debt?}

We observed that developers most frequently report \textit{Design Debt} (33/81), followed by \textit{Requirement Debt} (17/81) and \textit{Test Debt} (17/81). Compared to prior SATD distributions, design-related issues are proportionally lower, while requirement and testing debts are higher, indicating that AI-assisted development tends to shift self-admitted debt toward later development stages, particularly requirement completion and validation activities.

\noindent
{\bf RQ2: How do developers attribute the role of AI in generative-AI induced SATDs?}

We found that developers often attribute AI-assisted code as a trigger for potential issues, expressing uncertainty about its completeness and correctness and noting the need for more verification or revision (34/81 cases). 
In several instances, developers hold AI-assisted code responsible for errors, redundant logic, or unstable behavior that must be resolved later (22/81 cases).
At the same time, they also recognize AI assistance as beneficial for improving, refining, or validating existing implementations which contributes to resolving existing technical debts (19/81 cases).

Our findings show that AI-assisted development reshapes the distribution of technical debt, with fewer design-related issues but a higher prevalence of requirement and testing debts. Beyond these shifts, we identify a distinct form of debt is emerging when AI-generated code is adopted without full developer understanding or confidence. We term this \textbf{{\name} ({\nameacronym{}})}.

To the best of our knowledge, this is the first empirical study that systematically examines self-admitted technical debt explicitly linked to generative AI usage in source code comments. 
Looking ahead, our work posits several directions to examine how AI involvement influences the emergence and evolution of technical debt across the software development life cycle (SDLC), and how development tools or team practices might adapt to identify, document, and manage these AI-related debts more effectively.

\textbf{DATA AVAILABILITY} The dataset is available at ~\cite{Mujahid_2026_zenodo_18194031}. 

\section{Related Work}
\subsubsection*{Self-Admitted Technical Debt.}
Potdar and Shihab conducted an exploratory study on SATD by manually analyzing code comments to characterize how developers acknowledge and describe technical debt~\cite{potdar2014satd}. Maldonado et al. introduced a widely used taxonomy and later developed NLP-based methods for automatic SATD detection~\cite{maldonado2015detecting, maldonado2017nlp}. 
Li et al. conducted a systematic mapping study on technical debt and its management, categorizing major debt types, summarizing management strategies, and identifying gaps in empirical evidence and automation that shaped subsequent research on technical debt~\cite{li2015systematic}.
Large-scale empirical analyses further quantified SATD prevalence and types across open-source projects~\cite{bavota2016large,Huang2018SATD}. Subsequent work compared SATD practices in industry and open source~\cite{zampetti2017satd}, examined keyword-based detection~\cite{rantala2020prevalence,rantala2024keyword}, and proposed refined taxonomies~\cite{cassee2022self}. Other research extended detection beyond comments to multiple artifacts such as commits, issues, and pull requests~\cite{Li2023MultiSourceSATD}. In parallel, automation efforts have mapped tools for technical-debt management across the development lifecycle~\cite{biazotto2024technical}.

\subsubsection*{AI-Assisted Software Development.}
Recent studies have examined how generative AI is influencing software engineering practice. Sauvola et al. discussed how AI may reshape development processes and responsibilities~\cite{sauvola2024future}, while Coutinho et al. reported productivity gains alongside reliability and trust challenges~\cite{coutinho2024role}. Empirical studies of GitHub Copilot and other LLM tools have analyzed developer–AI collaboration and code quality~\cite{barke2023grounded,du2024evaluating,jin2024canchatgpt,guo2024exploring}. Related work has also explored AI for identifying or managing technical debt~\cite{melin2025exploring,li2025large}.

Despite these advances, little has been explored about how the use of AI itself contributes to the formation of technical debt. Our study addresses this gap by analyzing code comments that explicitly mention both AI involvement and self-admitted technical debt.

\section{Methodology}

\subsection{Data Collection}
\subsubsection{Collection of LLM Referenced Comments}
We collected source code comments from public GitHub repositories between November 2022 and July 2025, focusing on Python and JavaScript as these two are the most widely used languages in open-source ecosystem~\cite{so2024tech,octoverse2024blog,octoverse2023}. To identify comments where developers explicitly mention the use of AI/LLM tools, we built 196 structured search queries combining (A) 7 AI related terms (e.g., \textit{LLM, AI, GPT, ChatGPT, Copilot, Gemini, Claude}), (B) 6 generative verbs (e.g., \textit{generated, suggested, written}), and (C) 4 connector terms (\textit{by, from, with, using}).
Queries of the forms (A + B) and (B + C + A) were executed via the GitHub Code Search API~\cite{github_code_search_api}, yielding 37,234 files. We then used AST parsing to extract matched comments~\cite{tree_sitter}. Since many comments were duplicates (e.g., {\tt ``generated by ChatGPT''}, we removed them. After deduplication, we found 6,540 unique comments.

\subsubsection{Debt Acknowledgment Detection}
Within this LLM-referenced set, we applied a second filter using canonical \textit{self-admitted technical debt} keywords recognized in empirical software engineering research: \texttt{TODO, FIXME, HACK, XXX}~\cite{rantala2020prevalence}.
The search was performed using regular expressions with case-insensitive matching to capture variations such as \texttt{// todo} or \texttt{\# FixMe}.

\subsubsection{Resulting Dataset}
The intersection of these two filters yielded 96 unique comments that both (a) acknowledged the use or influence of AI/LLM and (b) contained explicit indicators of technical debt. Of the collected comments, 1.47\% met our criteria, which is close to the 1.86\% of SATD comments reported by~\cite{maldonado2017nlp}.

\subsection{SATD Types}

\subsubsection{Taxonomy of SATD}
\label{sec:taxonomy-satd}
To ensure compatibility with prior research on self-admitted technical debt, we adopted the widely used taxonomy by Maldonado et al.~\cite{maldonado2015detecting}. They proposed 5 common SATD types:

\begin{itemize}[leftmargin=*]
    \item \textbf{Design Debt}: These comments show design flaws in the code, such as misplaced logic, missing abstractions, overly long methods, or temporary workaround implementations.
    \item \textbf{Defect Debt}: Comments where author states that there is a defect in the code.
    \item \textbf{Documentation Debt}: Authors clearly mention the need for documentation of the code.
    \item \textbf{Requirement Debt}: Expresses incompleteness of the code or a unit of code such as: class, function or method.
    \item \textbf{Test Debt}: Shows the need for implementation or improvement of test.
\end{itemize}

\subsubsection{Annotation}
Two annotators independently labeled the 96 source code comments following the guidelines provided by Maldonado et al.~\cite{maldonado2015detecting}, as detailed in their replication package~\cite{maldonado2025_tseSatdData}. Each comment was classified into one of the five SATD categories described in Section~\ref{sec:taxonomy-satd}. Inter-annotator agreement, measured using Cohen’s~$\kappa$=0.896, indicated high consistency. Disagreements were resolved through in-person discussion.

During annotation, we found 10 instances where AI/LLM mention without actual usage of AI, and in 5 instances did not contain technical debt. These 15 instances were labeled as \textit{False Positive} and removed from the dataset, yielding a final set of 81 annotated comments. 

\begin{table*}[tb]
\centering
\caption{Distribution of Technical Debt Types}
\label{tab:td-types}
\begin{tabular}{p{0.15\textwidth} p{0.25\textwidth} p{0.4\textwidth} p{0.10\textwidth}}
\hline
\textbf{Debt Type} & \textbf{What the comment addresses} & \textbf{Example} & \textbf{Count} \\
\hline
Design Debt & Problematic design, misplaced code, poor implementation & 
\texttt{``TODO - this is copilot generated code, needs refactoring to a kdata object'' 
}
& 33 (40.74\%) \\ \hline

Requirement Debt  & Incomplete/partial implementation & ``{\tt TODO: Add parameter to include ingredients from the gpt generated check} [...]’’ & 17 (20.98\%) \\ \hline

Test Debt & Needs test/verification/validation of AI supported code & ``{\tt TODO: test this Copilot generated code’}'' & 17 (20.98\%) \\ \hline

Defect Debt & Bug/defect in code & ``{\tt TODO fix this ChatGPT created code.}'' & 11 (13.58\%) \\ \hline

Documentation Debt & Incomplete/improper documentation & ``{\tt TODO (USERNAME): This comment is generated by ChatGPT, which may not be accurate.}'' & 3 (3.70\%) \\
\hline
\textbf{Total} & & & \textbf{81} \\
\hline
\end{tabular}
\end{table*}

\subsection{Identifying the Role of AI}
To understand how developers attribute the role of AI in LLM-referenced SATD comments, we manually analyzed all 81 comments using an open coding approach, where categories and concepts were inductively derived from the data rather than predefined~\cite{strauss1990basics, seaman1999qualitative}.
As no prior taxonomy describes how AI contributes to technical debt, the two authors collaboratively developed through an iterative discussion. They reviewed the comments, compared interpretations, and refined emerging themes until stable and consistent categories were established. This process led to 4 types of roles:

\begin{itemize} [leftmargin=*]
    \item \textbf{Source}: 
    Developers indicate that AI-generated code directly introduces problems such as incorrect logic, incomplete implementation, redundant code, or temporary fixes. In these cases, AI is explicitly described as causing the debt. 
    Example: \texttt{``TODO:ChatGPT suggested super().close() and it crashed so I added the if. I don’t about this.''}
    
    \item \textbf{Catalyst}: 
    Developers express uncertainty about AI-generated contributions that function but may require later verification. AI does not cause an immediate issue but prompts awareness of potential future debt.  
    Example: \texttt{``TODO! validators generated by copilot , should be verified :works but doesn't mean it works all the time''}.
    \item \textbf{Mitigator}: Developers mention using AI to help address existing debt, such as by generating tests or refactoring suggestions.
    Example: \texttt{``TODO - Try these tests, generated by Copilot''}
    
    \item \textbf{Neutral}: AI or technical debt terms are mentioned without a clear link to the creation or resolution of technical debt.
\end{itemize}

The annotation instructions for open coding are available in our replication package~\cite{Mujahid_2026_zenodo_18194031}.

\section{RQ1: What types of self-admitted technical debt emerge in developer comments that acknowledge both LLM involvement and technical debt?}

We begin by examining the types of self-admitted technical debt that appear in comments explicitly mentioning LLM use.

\subsection{Results}
Table \ref{tab:td-types} summarizes the types of SATD identified in code comments that explicitly mention generative-AI usage. 
The most common category is \textit{Design Debt (33/81)}, which reflects issues such as misplaced code, poor implementation or a temporary workaround. 
For example, \texttt{``TODO - this is copilot generated code, needs refactoring to a kdata object’’} is indicating that the implementation needs to be refactored, adding a design debt to the developers.
Similarly, \texttt{``TODO: Modify this component to fit your needs. The ProfilePage component was generated with Github Copilot.’’} suggests that  the developers need to modify the AI-suggested component to fit project-specific needs. 

\textit{Requirement Debt} appears in 17/81 comments. 
Figure \ref{fig:req-comment} illustrates a typical example where a function was initially generated by Gemini but left unfinished. 
Another comment, \texttt{``TODO: Add parameter to include ingredients from the gpt generated check''}, demonstrates
a deferred integration task where AI generated code needs to include the additional parameters.

We observed \textit{Test Debt} in 17/81 cases, where developers defer testing or validation of the integrated AI-assisted code, e.g., ``{\tt TODO: test this Copilot generated code}''. 
There are also comments mentioning about tests generated by generative AI, which requires further improvement, e.g., \textit{``TODO 2023-08-23 10:41: - [ ] Skeleton of tests written by ChatGPT, write tests''}.

We further noticed \textit{Defect Debts} in 11/81 instances, where developers acknowledging a defect/bug that needs to be fixed. For example, comments such as {\tt ``TODO fix this ChatGPT created code.''} and {\tt ``TODO: Does not work. It's just generated from ChatGPT''} illustrate those. 
A very small portion were \textit{Documentation Debt (3/81)}, e.g., {\tt ``TODO (USERNAME): This comment is generated by ChatGPT, which may not be accurate.''}

\subsection{Discussion.}
\textit{Design Debt} appears most frequently, but its proportion is notably lower than in prior studies, while \textit{Requirement} and \textit{Test} Debts are comparatively higher. This shift suggests that AI-assisted development may influence how debt is introduced: developers rely on generative models to produce initial structures but defer completion and validation. The higher share of \textit{Test Debt} indicates that AI-generated code often enters projects without immediate verification, reflecting postponed quality assurance. 
Despite our smaller dataset size compared to Maldonado et al.~\cite{maldonado2017nlp,maldonado2025_tseSatdData}, our findings suggest a possible difference in the distribution of technical debt categories. In particular, we observe a lower proportion of design-related issues (40.74\% vs.\ 71.84\% in Maldonado et al.), alongside comparatively higher shares of requirement (20.98\% vs.\ 14.24\%) and testing-related concerns (20.98\% vs.\ 2.09\%). Taken together, these differences suggest a possible tendency for AI-generated code to be associated more often with implementation and evaluation activities than with upfront design decisions, although these observations remain exploratory given the dataset size, although further investigation with larger datasets is needed to substantiate this interpretation.

\section{RQ2: How do developers attribute the role of AI in generative-AI induced SATDs?}

While RQ1 established that AI-assisted development shifts the distribution of technical debt toward implementation and validation stages. However, these categories alone do not capture how developers perceive AI’s responsibility in creating or mitigating such debt. To address this, we examine the ways developers attribute roles to AI when discussing self-admitted technical debt.

\subsection{Results.}
\begin{figure}
    \centering
    \includegraphics[width=1\linewidth]{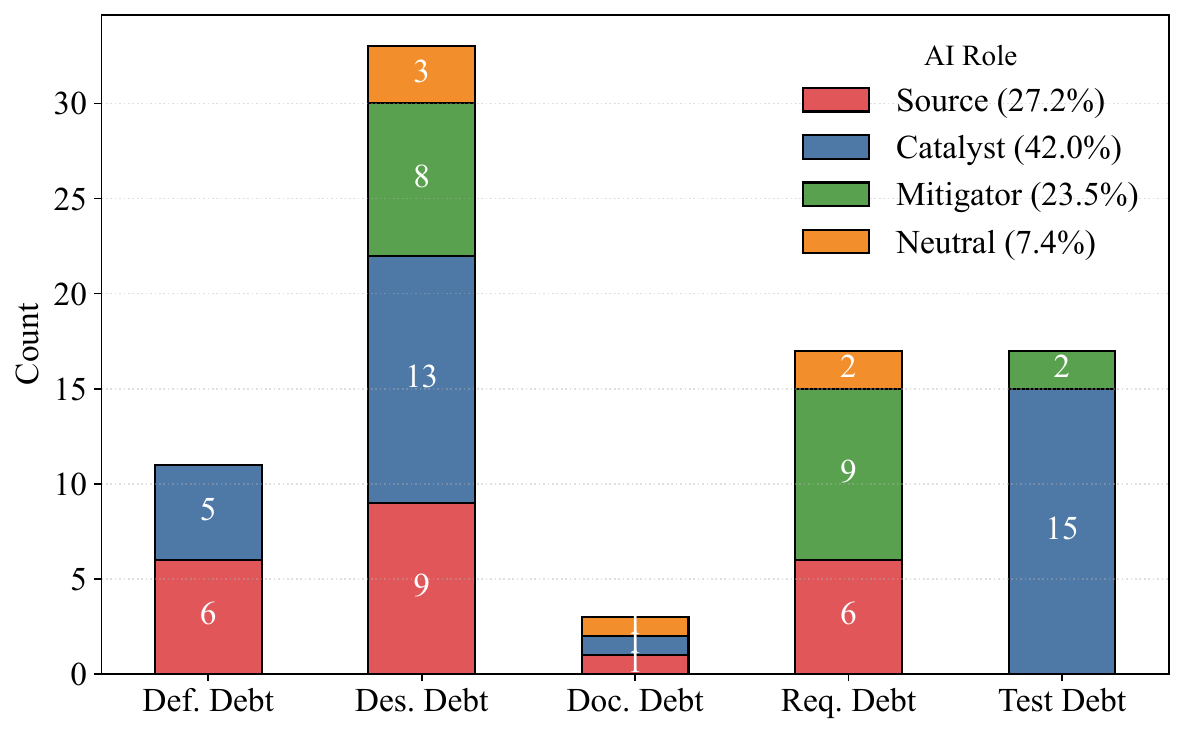}
    \caption{GenAI roles across SATD types}
    \label{fig:role-vs-type}
\end{figure}

Figure \ref{fig:role-vs-type} summarizes the roles developers attribute to AI within SATD-related comments. 
The most common role is that of a \textit{Catalyst - (34/81, 41.98\%)}, 
where AI-assisted codes prompt awareness of potential technical debt.
These comments express uncertainty about correctness or completeness, leading developers to defer testing and validation. 
For instance, \texttt{``TODO: generated by ChatGPT, don't know how reasonable this is''} shows doubt about the reliability of a generated artifact, while, 
{\tt ``TODO: AI Generated, please check the fields.''} records postponed checking. Here, AI functions as a \textit{Catalyst}, surfacing uncertainty that developers acknowledge as potential debt.

In 22/81 (27.2\%) cases, developers are identifying AI as a \textit{Source} of technical debt.
These comments describe situations where AI-generated code introduces errors, redundant logic, or unstable behavior that developers must later address. For example,
{\tt ``TODO: Does not work. It's just generated from ChatGPT''} signals nonfunctional output that requires fixing, while {\tt ``TODO: remove all unnecessary methods; this is an AI-generated file''}
points to redundant code. 

A third pattern, we observed in 19/81 cases, developers frame AI as a \textit{Mitigator}. Here, AI contributes to reducing existing debt by generating refactors, test scaffolds, or suggesting design alternatives. 
For instance, {\tt ``PJB replaced with `new' in each case. TODO: use an improved const as suggested by Claude''} here, Claude is suggesting how to mitigate a past debt
Finally, In 6/81 (7.4\%) of the comments are classified as neutral, where AI is mentioned but its role to technical debt cannot be clearly determined.


\medskip
\noindent
\textbf{Distribution of AI Roles Across SATD Types.}
Figure~\ref{fig:role-vs-type} shows how AI roles vary across SATD types. When AI is identified as the \textit{Source} of debt, it is most often associated with \textit{Design Debt} (9 instances), followed by \textit{Requirement Debt} and \textit{Test Debt} (6 instances each), and \textit{Documentation Debt} (1 instance).

As a \textit{Catalyst}, AI is primarily linked to \textit{Test Debt} (15 instances) and \textit{Design Debt} (13 instances). When acting as a \textit{Mitigator}, it is most commonly associated with \textit{Requirement }Debt (9 instances) and \textit{Design Debt} (8 instances), reflecting its use in supporting code generation and refactoring.

\subsection{Discussion.}
The most frequent role developers attribute to AI is that of a \textit{Catalyst}, suggesting that AI raises awareness of potential issues. 
Developers often treat AI-generated code as provisional material that may require additional testing or refactoring. 
This reflects a human-in-the-loop workflow in which AI speeds up implementation while developers retain responsibility for quality assurance. When AI is identified as the \textit{Source}, it is typically blamed for introducing faulty or redundant code linked to incomplete or unstable functionality. AI also appears as a \textit{Mitigator}, assisting in debt reduction by suggesting improvements, generating tests, or supporting refactoring, particularly in \textit{Design} and \textit{Requirement} Debt. 
Developers thus engage with AI in multiple, sometimes contrasting ways, as a potential risk, an opportunity for inspection, and a mechanism for improvement.

\section{{\name{}} ({\nameacronym{}})}
We observed a recurring pattern that developers are integrating AI-assisted code into their production codebases despite i) uncertainty about its correctness, and ii) limited understanding of its internal logic. 
We introduce this notion as a \textit{conceptual interpretation grounded in observed self-admitted comments}. This pattern appears across multiple SATD types discussed in Section~\ref{sec:taxonomy-satd}. 
For example, in the comment, \texttt{``TODO: Copilot suggested this function (I have no clue what the regex is doing)''}, the developer acknowledges incorporating an AI-generated function without understanding its behavior. 
Similarly, \texttt{``TODO: This is totally GPT generated and I’m not sure it works''} expresses uncertainty about correctness, and \texttt{``TODO: generated by ChatGPT, don’t know how reasonable this is''} reflects doubt about the plausibility of the AI output. 
Collectively, these comments illustrate a recurring form of uncertainty-driven technical debt that we refer to as \textbf{{\name} ({\nameacronym{}})}.

The observed pattern is consistent with prior work on automation bias, where developers rely on automated suggestions despite uncertainty about their correctness or rationale~\cite{rastogi2022deciding, agudo2024impact}. Research in human-AI interaction has shown that such reliance can influence judgment, verification behavior, and trust in automated systems~\cite{wingerter2025mitigating, rastogi2022deciding, agudo2024impact}. In our study, we observe how automation-related uncertainty is documented as self-admitted technical debt in source code comments.

{\nameacronym{}} characterizes situations in which uncertainty about AI-generated code is not immediately resolved but instead deferred, creating a latent burden for future development and maintenance. This form of debt manifests through two recurring dimensions:
\begin{itemize}[leftmargin=*]
    \item \textbf{Knowledge Deficit and Deferred Quality Assurance}: Developers integrate AI-generated code without a full understanding of its underlying logic or correctness, creating a gap between the artifact and their mental model of the system. This partial comprehension often leads to postponed validation or testing, as developers defer closer examination to a later stage. Such deferral constitutes a cognitive form of technical debt: understanding is temporarily outsourced to the AI, leaving uncertainty that may resurface during maintenance or modification.
    
    \item \textbf{Lack of Trust and Delegated Responsibility}: Developers’ uncertainty about the reliability or rationale of AI-generated code may erode their confidence in the system as a whole. To manage this uncertainty, responsibility for verifying or improving the AI’s output is implicitly shifted to other team members or future revisions. This delegation reflects an operational dimension of \textbf{{\nameacronym{}}}, where provisional integration of AI output allows unverified code to persist, creating ambiguity around ownership and long-term accountability.  
\end{itemize}
\section{Implications and Reflections}





\subsection{Technical Debt in AI-Assisted Development}
Our findings reveal a shift in the composition of SATD. \textit{Design} Debt has become less dominant, while \textit{Requirement} and \textit{Test} Debts are increasingly common.
The changing distribution of debt types suggests that AI assistance is reshaping not the taxonomy of technical debt but its underlying causes. In traditional settings, debt often results from conscious trade-offs or time pressure. In contrast, AI-assisted debt tends to emerge from partial reliance on automatically generated code and deferred human oversight. This shift implies that technical debt frameworks may need to account for delegated decision-making, where responsibility for quality and completion is shared between humans and AI systems. More broadly, our observations suggest that some AI-related debt may arise unintentionally, not from deliberate shortcuts but from uncertainty about the behavior or suitability of AI-generated code.

\subsection{Interpreting {\nameacronym{}}}
The notion of \textbf{{\nameacronym{}}} highlights how limited understanding of AI-generated code can affect long-term maintainability. When code functions without a clear rationale, comprehension and accountability may become fragmented, increasing future effort to modify, validate, or extend the code. In the analyzed comments, developers explicitly acknowledge such uncertainty when integrating AI-generated artifacts, indicating awareness of potential future maintenance challenges.

Integrating principles from \textit{explainable AI} and \textit{responsible AI} could help mitigate these risks by promoting transparency about how models produce code, documenting decision rationales, and clarifying accountability when AI assistance is used. Such practices may help developers retain insight into AI-generated code and better recognize when limited understanding could introduce additional maintenance effort over time.

\subsection{Integrating AI Practices into the SDLC}
The observed roles of AI, as Source, Catalyst, and Mitigator, point to the need for clearer integration within the software development life cycle (SDLC). Differentiating these roles can guide appropriate oversight, from validation and testing to documentation and provenance tracking. Such awareness may help teams balance AI’s supportive functions with the risks of introducing uncertainty or incomplete understanding into codebases.
\section{Threats to Validity}

This study relies on keyword-based detection of LLM mentions and self-admitted technical debt indicators, which may result in false positives and false negatives. Developers may reference AI-generated code using alternative phrasing not captured by our queries, and not all instances of \texttt{TODO} or \texttt{FIXME} necessarily indicate genuine technical debt. Although manual annotation was used to filter irrelevant cases, some subjectivity in interpretation remains.

The analysis is based on 81 AI-related SATD instances drawn from public Python and JavaScript repositories between November 2022 and July 2025. While this enables focused qualitative analysis, the limited sample size constrains quantitative generalization. The findings should therefore be interpreted as exploratory and may not generalize to proprietary projects, other languages, or future AI-assisted development practices as generative AI tools and developer workflows continue to evolve.

\section{Conclusion and Future Work}
This study provides an empirical examination of self-admitted technical debt that explicitly references the use of generative AI. By analyzing 81 LLM-related code comments, we observe that AI assistance is associated with a shift in how debt is self-admitted: design-related issues appear less frequently, while requirement and testing debts are more common, reflecting deferred completion and validation of AI-generated code. We propose {\bf {\name} ({\nameacronym{}})} as a conceptual lens to describe recurring patterns of self-admitted technical debt in which developers explicitly express uncertainty about the behavior or rationale of AI-generated code, inviting researchers to further examine, refine, and empirically validate this notion across broader contexts.

Our study represents an initial step toward understanding AI-related technical debt and how generative AI is influencing software development life cycle practices. The analysis is limited to Python and JavaScript projects; future work should extend to additional languages and ecosystems to assess whether similar patterns hold. Longitudinal repository studies and automated analysis techniques could further illuminate how AI-related debts evolve over time and whether they are resolved or accumulate as systems mature. Future research may also explore practices that improve the transparency and verifiability of AI-generated code, supporting developers in managing uncertainty and sustaining long-term maintainability in AI-assisted software development.

\end{sloppypar}

\balance
\bibliographystyle{ACM-Reference-Format}  
\bibliography{references}

\end{document}